\documentclass[preprint,proceedings]{rmaa}

\usepackage{paralist}

\usepackage{psfrag,color}
\usepackage{epsfig, psfig, epsf}

\SetYear{2003}
\SetConfTitle{Compact Binaries in the Galaxy and beyond}

\title{Simultaneous INTEGRAL/RXTE Observations of GRS 1915+105} 

\author{
  J. Rodriguez,\altaffilmark{1,2},
  D. Hannikainen,\altaffilmark{3}
  O. Vilhu,\altaffilmark{3,2}
  Y. Fuchs\altaffilmark{1}
  and S.E. Shaw\altaffilmark{4,2}}

\altaffiltext{1}{CEA/Service d'Astrophysique, Saclay, France.}
\altaffiltext{2}{Integral Science Data Center, Versoix, Switzerland.}
\altaffiltext{3}{University of Helsinki, Helsinki, Finland}
\altaffiltext{4}{University of Southampton, Southampton, UK}

\shortauthor{Rodriguez et al.}
\shorttitle{INTEGRAL/RXTE observations of GRS 1915+105}

\listofauthors{J. Rodriguez, D.C. Hannikainen, O. Vilhu, Y. Fuchs}
\indexauthor{Rodriguez, J.}
\indexauthor{Hannikainen, D.C.}
\indexauthor{Vilhu, O.}
\indexauthor{Fuchs, Y.}
\suppressfulladdresses

\abstract{We present the first results of simultaneous INTEGRAL
and RXTE observations of the microquasar GRS~1915+105.
We focus on the analysis of the unique highly variable observation and
show that we might have observed a new class of variability. We
then study the energetic dependence of a low frequency QPO from our
steady observations. }

\addkeyword{X-ray: observations}
\addkeyword{Stars: Binaries}
\addkeyword{Stars: GRS1915+105}

\begin{document}
\maketitle

\section{Generalities on GRS 1915+105}
\label{sec:intro}
\begin{figure}[htbp]
\centering
\epsfig{file=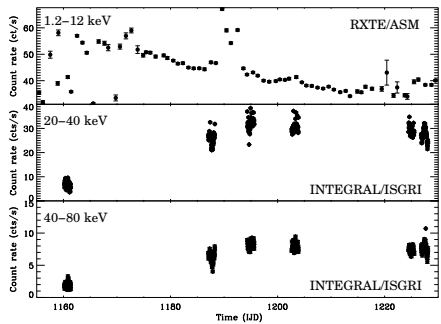,width=6cm}
\caption{RXTE/ASM (top) and INTEGRAL/ISGRI (middle and bottom) 
 light curves during spring 2003.}
\label{fig:lightcurve}
\end{figure}
Since its discovery in  1992 (Castro-Tirado et al. 1992), the study 
of GRS~1915+105 at all wavelengths  has given us a unique insight 
into the accretion-ejection coupling in microquasars. While the almost 
daily coverage with RXTE allowed to classify the X-ray variability 
 into 12 classes  (Belloni et al. 2000), multiwavelength 
coverage showed a correlation between radio ejections and X-ray 
variability (e.g. Mirabel et al. 1998).  GRS~1915+105, is 
also one of the 2 microquasars for which a compact jet has been imaged 
using radio interferometry (Dhawan et al. 2000, Fuchs et al. 2003).\\
\indent GRS~1915+105  was observed with INTEGRAL during 6 observations
of at least 100 ks (revolution 48, 59, 69, 70 + 57 \& 62 discussed
in Fuchs et al. 2003 and these proceedings).  
Simultaneous RXTE observations were performed 
during 4 of these 6 observations. Fig. \ref{fig:lightcurve} 
shows the RXTE/ASM daily light curve 
and and the INTEGRAL/ISGRI 
light curves over the 6 observations.  
\section{First INTEGRAL observation of GRS~1915+105}
\begin{figure}[htbp]
\begin{tabular}{ll}
{\hspace*{-0.6cm}\epsfig{file=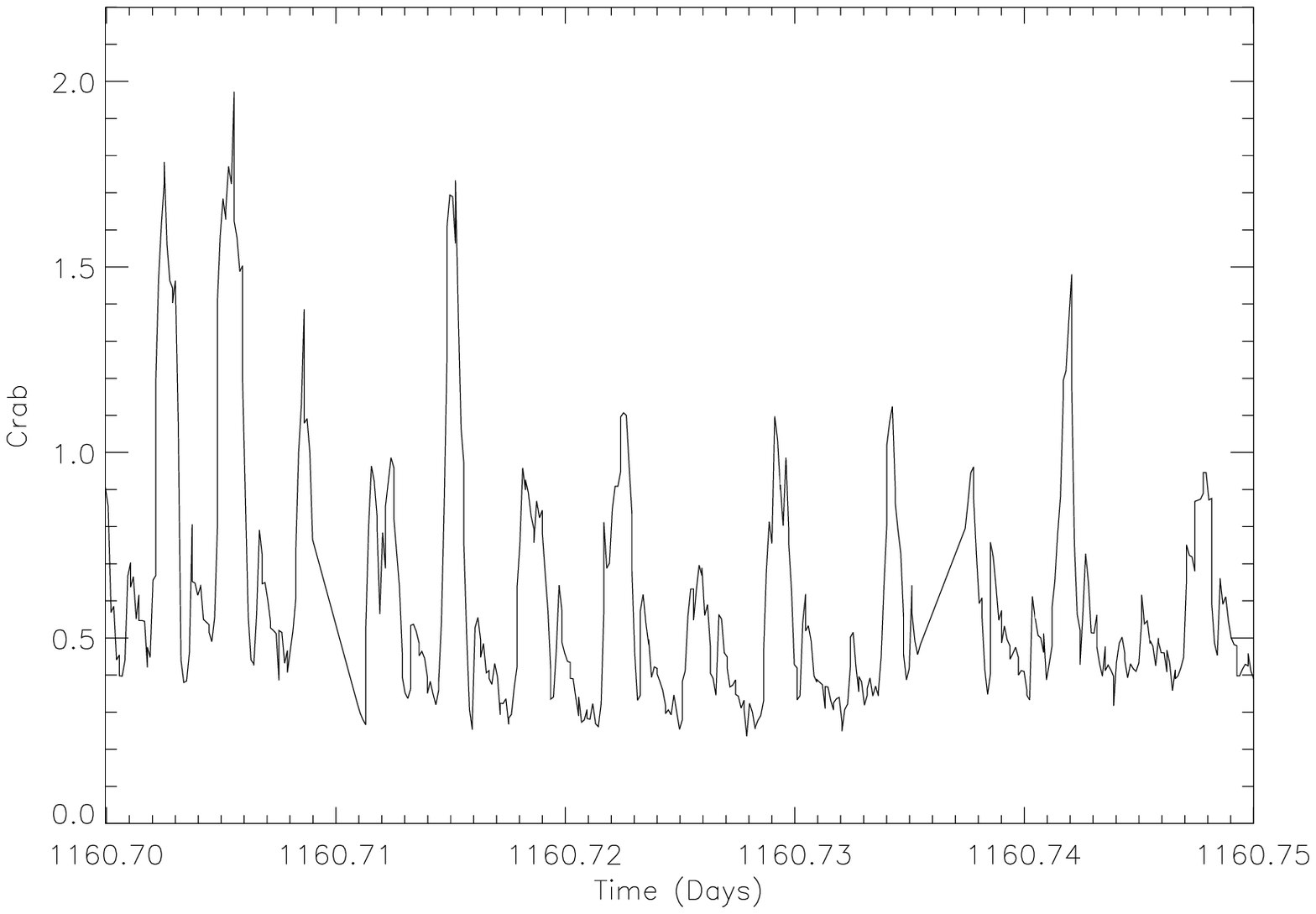,width=4.5cm}}&
{\hspace*{-0.6cm}\epsfig{file=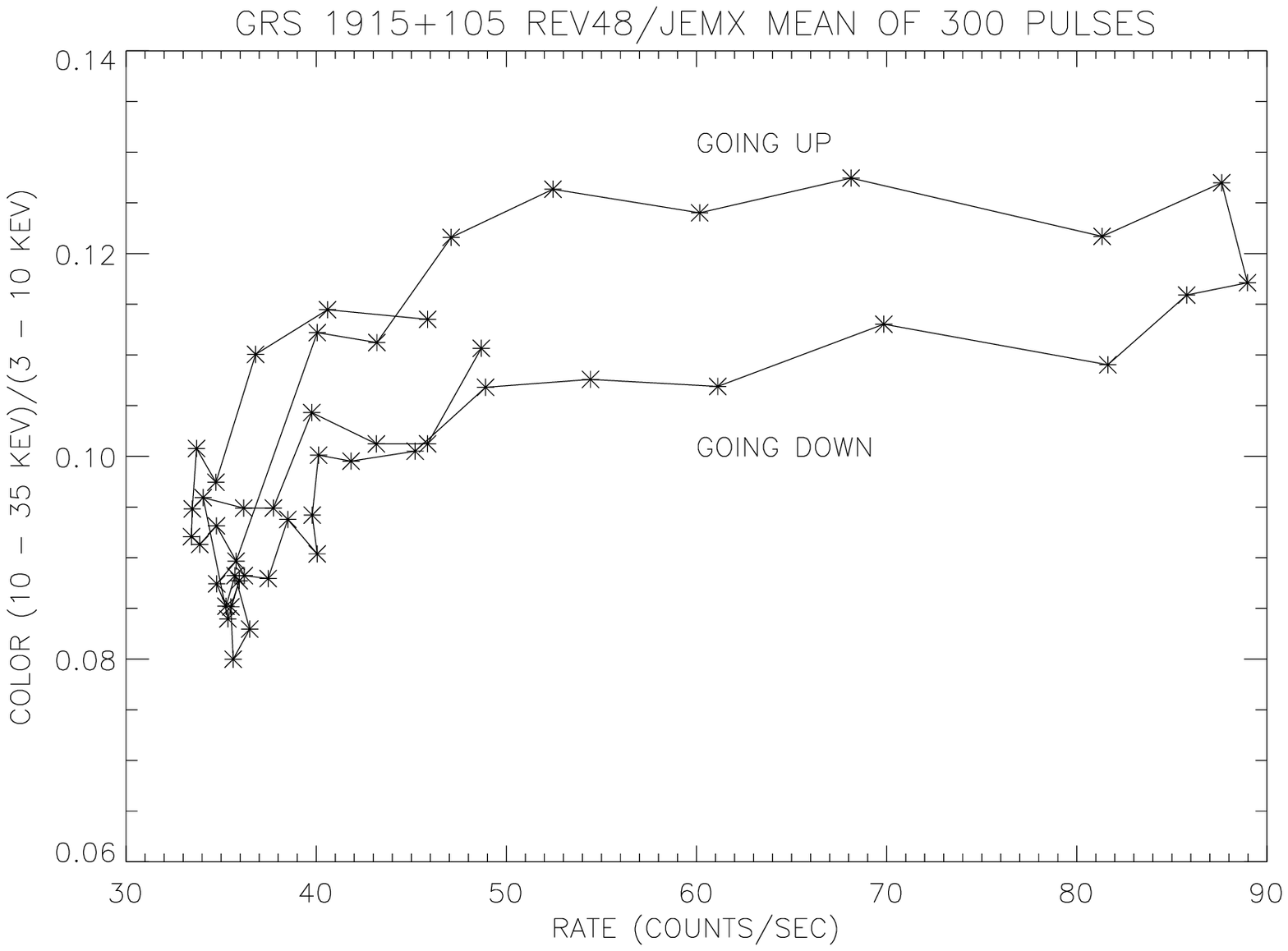,width=4.5cm}}\\
\end{tabular}
\caption{Left: Zoom on the JEM-X lightcurve during Rev. 48. Right: Mean hardness ratio vs. count rate, for 300 spikes as seen by JEM-X}
\label{fig:jemx48}
\end{figure}

The first observation  
shows the recurrence of large spikes on time scale of $\sim$5 min. (Fig. 
\ref{fig:jemx48}). The
production of a power spectrum showed a strong QPO at 3 mHz, indicating
that the observation is dominated by these variations (Hannikainen
et al. 2003). 
This light curve resembles the $\rho$ class of 
Belloni et al. (2000), however the time scale is different. A more 
precise study allowed us to further distinguish it from any 
class of variability known up to now. Indeed when we study the
hardness ratio of the source, it appears that the rising part of the 
spikes is hard and the decay is soft. This is not expected if 
we suppose the rising part of the spikes is due to a disk instability, 
and the decay represents the receding of the disk. This kind of 
behavior could be evidence
for the presence of two independent accretion flows, obeying different
time scales. In that case the disk would react to any perturbation of 
the accretion rate on a viscous time scale, while the corona would react
on a free fall time scale. Such a conclusion has been drawn in the case 
of other black hole binaries (with low mass stars), e.g. 
1E 1740.7--2942, GX 339--4, GRS 1758--258 (Smith et al. 
2002) or XTE J1550--564 (Rodriguez et al. 2003). 
The recurrence time of the peaks in GRS~1915+105 is not comparable with 
any of these black hole binaries, however. More precise studies of this 
class, e.g. time resolved spectroscopy and comparison between the
dips, spikes, rising and decaying phases of the peaks should allow
us to obtain more information on the physics underlying these 
phenomena.
\section{Steady state observations: QPO studies}
\begin{figure}[htbp]
\centering
\epsfig{file=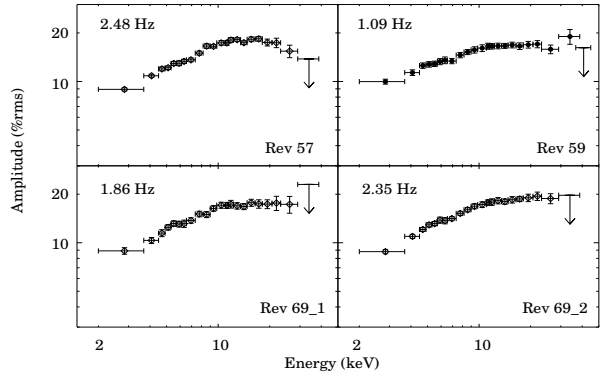,width=\columnwidth,height=5cm}
\caption{Energy dependence of the LFQPO detected during the 4 observations 
labeled in each panel. The frequency of the feature is also indicated.}
\label{fig:qpovskev}
\end{figure}

GRS~1915+105 was observed in a steady  state during all the 
other observations. A typical example of RXTE/INTEGRAL spectral analysis
is presented in Fuchs et al. (2003, and these proceedings). During
 all the observations for which we had simultaneous RXTE coverage,
 the spectral parameters are comparable. In all observations a
strong ($\sim14\%$ rms amplitude) low frequency quasi-periodic 
oscillation (LFQPO) is clearly detected (e.g. Fuchs et al. 2003). 
Given the long RXTE exposure times and the high luminosity of the 
source we could study the energetic dependence of the LFQPO parameters 
with the highest possible spectral resolution. Both the frequency and 
the width do not depend on the energy. The evolution of the amplitude
vs. the frequency is plotted in Fig. \ref{fig:qpovskev} for rev 57, 59
and 69.  
A cut-off around 15 keV is clearly detected in Rev. 57, 
confirming previous findings (e.g. Rodriguez et al. 2002). The position of 
the cut-off might be simply shifted towards high energies in the other 
observations.
The statistics of the PCA data above 25 keV prevents, however, any firm 
conclusions. The origin of the cut-off is still unclear, and might indicate 
different phenomena: either the QPO moves relative to the corona  
(see e.g. Tomsick \& Kaaret 2001), or the shape of the QPO spectrum might be 
evidence for a hot spot rotating in the disk (Rodriguez et al. 2002).
In a simpler manner, if a significant part of the high energy photons 
is emitted through synchrotron radiation by a compact jet (which is 
present at least during rev 57; Fuchs et al. 2003), then the shape of 
the QPO spectra can be easily understood if one assumes the QPO 
does not originate in the jet. Once again this would point toward a strong 
coupling 
of the QPO with the  Compton corona as usually observed, and in agreement
with most models of LFQPOs (e.g. Accretion Ejection Instability, or 
CENtrifugal Dominated BOundary Layer). In the future the use of INTEGRAL 
will help to better constrain  the spectra of QPO at high energy, and probe 
their relation to the spectral parameters while performing wide band 
spectroscopy.

\acknowledgements
The authors thank T. Belloni, M. Tagger, P. Varni\`ere, G. Henry, P.-O. Petrucci, C. Cabanac, S. Chaty, M. Ribo, N.J. Westergaard for useful discussions and comments on the present work. JR and YF acknowledge financial support from the CNES. DH is a Research Fellow of the 
Finnish Academy.

\end{document}